\documentclass[
aps,prl,twocolumn,
groupedaddress,superscriptaddress,
amsmath,floatfix,
showpacs
]{revtex4-1}

\usepackage{graphicx}  
\usepackage{bm}        
\begin{document}

\title{Channeling of Electrons in a Crossed Laser Field}

\author{S. B.~Dabagov} 
\email[]{sultan.dabagov@lnf.infn.it}
\affiliation{INFN LNF, Frascati (Roma), Italy} 
\affiliation{LPI RAS, Moscow, Russia}
\affiliation{NRNU MEPhI, Moscow, Russia}
\author{A. V.~Dik} 
\affiliation{LPI RAS, Moscow, Russia}
\author{E. N.~Frolov} 
\affiliation{LPI RAS, Moscow, Russia}
\affiliation{NR TPU, Tomsk, Russia}

\date{\today}

\begin{abstract}
  In this article a new analytical description of the effective interaction
  potential for a charged particle in the field of two interfering laser beams
  is presented. The potential dependence on the lasers intensities, orientation
  and parameters of the particle entering the considered system is analyzed. It
  is shown for the case of arbitrary lasers crossing angle that for different
  values of projectile velocity the attracting potential becomes a scattering
  one so that the channel axes and borders interchange each other. In addition
  the projectile radiation spectral distribution is given and general
  estimations on the expected beam radiation yield are outlined.
\end{abstract}

\pacs{61.85+p, 41.75.Ht, 41.90+e, 61.80.Fe}

\maketitle

\section{Introduction}
The topic of electrons dynamics in crossed laser beams is gaining growing
attention. Usually the case of standing electromagnetic wave being the result of
two counter-propagating laser beams is considered. Kapitza and Dirac were the
first who referred to electrons dynamics in such an interference field in well
known paper \cite{KapDir33} introducing Kapitza-Dirac phenomenon.  They have
described for the first time the possibility of electron beam diffraction in
standing optical field. Nowadays electrons diffraction on a standing
electromagnetic wave is used e.g. in Shintake monitor \cite{shint92} for beam
diagnostics.

Moreover, many papers were published lately proposing new-type free electron
laser with optical undulator \cite{fed88,bal10,andepj11,andprl12,andjop13},
channeling radiation source \cite{Bert86,Akh91,rreps} (all being based on
electrons dynamics in the field of two laser beams) and works clarifying
processes in such a system \cite{kaplanprl,kaplanpra,smorenburg11}. In this work
the electron motion in optical lattice formed by crossed linearly polarized
laser beams, in plane electromagnetic waves approximation, is described in terms
of particles channeling.

\subsection{Channeling phenomenon}
Usually channeling phenomenon is related to the beam propagation in aligned
crystals. As known the beam channeling in crystals takes place during
relativistic charged particles motion in periodic structures of the crystals
close to the main crystallographic directions that form the crystal axes or
planes. For relativistic electron traveling almost parallel to that directions
the potential of interaction between the electron and a set of the lattice atoms
(ions) could be averaged along the propagation direction. Potential well formed
in such a way can limit transverse motion of the projectile within well defined
channels, i.e. relativistic particle becomes undulating in transverse plane at
fast longitudinal motion down to the channel \cite{Lin65, Gem74}. For more than
50 years of intense studies the basics of crystal channeling for charged beams
have been in details defined and described in many scientific papers and books,
discussed in a number of conferences and workshops. Presently crystal channeling
is known as a useful technique for beam steering, while the related phenomena to
crystal channeling are promising candidates for coherent radiation sources (for
details, see in \cite{Kum76}). Moreover, the phenomenology of beams channeling
becomes very useful for describing neutral beams handling with help of various
beam guiding structures \cite{Buk06,DabUFN03}. Besides, channeling conditions
could be realized for particles not only in medium (crystals, capillaries
\cite{Buk06,DabUFN03} and nanotubes \cite{Zhev03,Klim96,Karab13}, plasma
\cite{es02,pu03,mal10,dnimb}) but also in high intensity electromagnetic fields
of specific configurations \cite{fnimb} that is the scope of this
article. Generally saying, channeling phenomenology may be applied for any kind
of charged or neutral particle beams motion in the external fields defined by
long transversely limited channels. And various features of the structure as a
periodic structure, for instance, may supply additional peculiarities of beam
passing through such structures.

\subsection{Ponderomotive potential}
In the region of two laser beams overlapping the ponderomotive force
characterized by averaged effective potential affects charged particles. The
ponderomotive potential forms planar (one-dimensional) potential wells. It was
shown that a charged particle can be trapped in such a well and oscillate in
it. The typically used ponderomotive force expression \cite{gaponov,bolot} is
written in the following way
\begin{equation}
  \label{eq:pond1}
  \mathbf{F}_p = \frac{e^2}{4\omega_0^2m}\vec{\nabla}\left|\mathbf{E}(x)\right|^2,
\end{equation}
where $e$ and $m$ are, correspondingly, the charge and mass of the electron,
$\omega_0$ is the laser frequency and $\mathbf{E}(x)$ is the electrical field
amplitude distribution over $x$-coordinate.

Lately Kaplan {\it et~al} have pointed out that the ponderomotive force
expression is much more complex \cite{kaplanprl,kaplanpra}. This force depends
not only on lasers intensities but also on their polarization and energy of the
particle. One of the most interesting results in their work is ponderomotive
potential inversion description. In brief, if the longitudinal (parallel to the
channels axes) speed of a charged particle placed in the field of two counter
propagating laser beams equals some defined critical value then the particle
feels no periodic ponderomotive potential at all, if the beams are
p-polarized. In other words, the height of one-dimensional potential channels
borders becomes zero. And if the particle speed is greater then the critical
value, ponderomotive potential sign inversion is observed. That means,
attracting potential regions become scattering ones and vice versa. Apart of
this peculiarity, the motion of a charged particle in presence of two
counter-propagating laser beams could be described as a sum of slow averaged
motion in the effective ponderomotive potential and simultaneously rapid small
oscillations defined by the lasers frequency. Such motion is treated by
different authors as betatron oscillations \cite{and13}, FEL oscillations
\cite{fed88,bal10,andepj11,andprl12,andjop13}, channeling oscillations
\cite{Bert86,fed88,Akh91,Art07,ihce} or simply called averaged (slow)
oscillations \cite{KapDir33,kaplanprl,kaplanpra,smorenburg11}. We here shortly
outline why channeling analogy is chosen by us.
\subsection{Optical lattice channeling}
As known, there are much in common between particles dynamics in crossed laser
beams and processes found in FEL undulators, betatron oscillations in plasma
channels and crystal channeling. We should give a short remark on why the
channeling point of view could be the most appropriate for considering the
phenomenon.

First of all, interference of two crossed laser beams creates electromagnetic
field peaks and nodes, i.e. optical lattice, which is similar to the crystal
lattice. This creates semblance of crystal lattice in absence of actual
medium. Furthermore, averaging interaction of a particle with both crystal and
optical lattices one derives effective potential responsible for particle
channeling in these systems. And this descriptive similarity is not the only
reason for treating the considered process as channeling.

Another reason lies in the similarity of possible applications and effects found
in optical lattice and in crystal channeling. As shown in
\cite{fed88,Art07,rreps,Akh91,Bert86} and described below, the effective
potential of both crystal and optical lattices could be very similar and both
are capable of trapping electrons. So that channeled electron beams may be
transported, focused and reflected by potential of both lattices. Bending a
crystal one obtains a tool for charged particles beams steering \cite{tsy79} and
bent laser channels may be also applied for this. Such channels may be formed by
illuminating a curved reflecting surface with a laser at some angle that creates
interference region with potential channels near the surface. The peculiarities
of beam channeling in a bent laser field will be described in a separate work.

Besides, recently various researches have shown different regimes of charged
particles dynamics in presence of intense laser fields that are analogous to
crystal volume reflection, to both planar
\cite{rreps,Akh91,Art07,Bert86,Bert87,fnimb} and axial \cite{Akh91} crystal
channeling and optical undulating.
\section{Electrons in crossed laser field}
The results below were derived for a relativistic electron in the field of
crossed p-polarized laser beams using plane wave approximation. The averaged
field in such a system is characterized by the planes of equal potential forming
planar channels. We show here channels parameters dependence on the lasers
crossing angle.

\begin{figure}
\includegraphics[width=5cm]{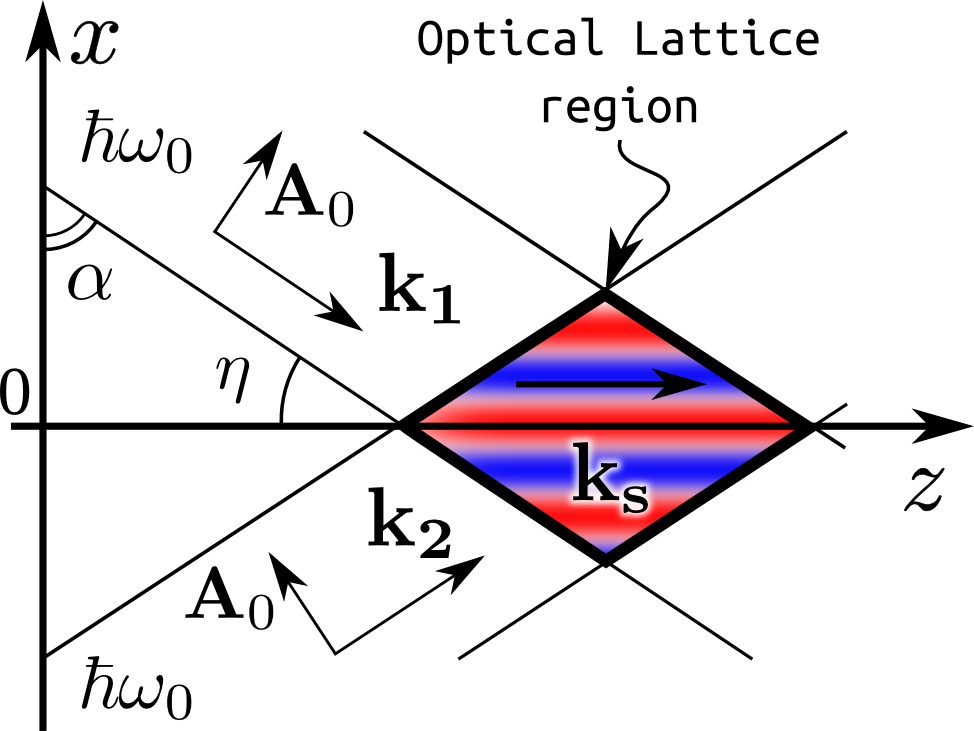}
\caption{\label{f1} The general scheme of the considered system. The effective
  potential channels (shown in color) axes are parallel to $0z$-axis.}
\end{figure} 

\subsection{Interaction potential}
Let us consider two crossed laser beams of the $\omega_0$ frequency with the
wave vectors $\mathbf{k_1}$ and $\mathbf{k_2}$, which lie in the $xOz$ plane at
the acute angles $\eta$ with respect to the $Oz$ direction (see Fig.~\ref{f1}).
Let the electron move in the combined laser filed, which is situated in the
region where two laser beams form a standing wave structure. Using plane wave
approximation, the electromagnetic field in some region of beams overlapping
could be described by a set of equations
\begin{equation}\label{field}
\left\{
   \begin{array}{c}
     \varphi \\
     A_x \\
     A_y \\
     A_z
   \end{array}
\right\} = 
\left\{
  \begin{array}{c}
    0 \\
    -2 A_0 \sin\alpha\cos(k_x x) \cos(\omega_0 t - k_z z) \\
    0 \\
    2 A_0 \cos\alpha\sin(k_x x) \sin(\omega_0 t - k_z z)
  \end{array}\right\}\,,
\end{equation}
where $k_{x,z}$ are the absolute values of appropriate $\mathbf{k}_{1,2}$
components, and $A_0$ is the longitudinal vector potential amplitude.

This structure is similar in some way to the crystal lattice, seen by a
high-velocity projectile, that is a set of crystal nodes, forming
crystallographic planes and axes. The standing waves can be represented as the
effective potential wells, the channels, periodically situated in a transverse
plane that can trap the electron at specific conditions. The non-relativistic
electron trajectory can be classically derived with the help of the well known
Kapitza formalism \cite{LandMech1988}, within which the trajectory of a
channeled electron could be presented as a sum of slow channeling oscillations
and rapid small amplitude oscillations. Such approach allows one to write down
the analytical expression for channeled electron trajectory in the case of small
channeling oscillations near the channel bottom. For the case of large amplitude
channeling oscillations the motion equation can not be linearized. Thus, only
qualitative estimations and computer simulations are feasible.
\subsection{Relativistic factor}
It should be underlined that to use Kapitza method becomes impossible directly
for relativistic electrons since their speed is comparable with the speed of
light. However, one may switch to new variables
\begin{equation}\label{eq1}
  S=(z-v_0t)P_z';\, 
  \mathbf{r}_{\bot}'=\mathbf{r}_{\bot};\, 
  \mathbf{P}_{\bot}'=\mathbf{P}_{\bot};\, 
  z'=z-v_0t,
\end{equation}
where $v_0$ is the electron initial longitudinal speed ($v_0\to c$), $P_i$ is
the generalized momentum projection on the $i^{th}$ axis. Thus, for the electron
in the field characterized by the vector-potential $\mathbf{A}$ the motion
equations are
\begin{equation}\label{eq2}
  \dot{p}_i' =-\frac{e}{c}\left[\frac{\partial A_i}{\partial t}
  -v_0\frac{\partial A_z}{\partial x_i'}
  +\dot{x}_j'\left(\frac{\partial A_i}{\partial x_j'}
  -\frac{\partial A_j}{\partial x_i'}\right)\right]
\end{equation}
\begin{equation}\label{eq3}
  \dot{x}_i'=\frac{cp_i'}{\sqrt{(mc)^2+p_j'p_j'}}-v_0\delta_{iz},
\end{equation}
where $p'_i$ is the kinematic momentum, $\delta_{ij}$ is delta Kronecker
symbol. This equations could be solved within Kapitza method, i.e. the electron
trajectory as mentioned above may be expressed as a sum of slow and rapid
oscillations $x'_i=\bar{x}_i+\xi_i$. The same is true for electron momentum
$p'_i=\bar{p}_i+p^\xi_i$. The right hand side of Eq.~(\ref{eq2}) then may be
written in the following way
$F_i(\mathbf{x}',v_0,t)+\dot{x}_j'B_i^j(\mathbf{x}',t)$, thus~-- taking into
account that
$\mid\dot{\mathbf{x}}'/c\mid\ll 1;~p_z'\gg \mid\mathbf{p}_{\bot}\mid'$~-- the
motion equations are spitted for slow oscillations
\begin{equation}\label{eq5}
  \left\{
      \begin{array}{ll}
        \dot{\bar{p}}_i=
        \overline{\xi_j\dfrac{\partial F_i(\bar{\mathbf{x}},v_0,t)}{\partial\bar{x}_j}}
        +\overline{\dot{\xi}_jB^j_i(\bar{\mathbf{x}},t)}, \\
        \bar{\mathbf{p}}_{i}=\bar{\gamma} m\dot{\bar{x}}_{i},
      \end{array}
    \right.
\end{equation}
and rapid oscillations
\begin{equation}\label{eq4}
  \left\{
    \begin{array}{ll}
      \dot{p}_i^{\xi}=F_i(\bar{\mathbf{x}},v_0,t), \\
      p_z^{\xi}=\bar{\gamma}^3m\dot{\xi}_z;\quad
      \mathbf{p}_{\bot}^{\xi}=
      \bar{\gamma} m\dot{\xi}_{\bot}\mathbf{e}_{\bot},
    \end{array}
  \right.
\end{equation}
where $\bar{\gamma}=\sqrt{1+\left({\bar{p}_z}/{mc}\right)^2}$.  The averaged
longitudinal impulse in the system is constant
\begin{equation}\label{eq9}
  \bar{p}_z=\frac{mv_0}{\sqrt{1-\beta_\parallel^2}},
\end{equation}
Averaging by rapid oscillations one derives the effective potential expression
\begin{equation}
  \label{ueff1}
  U_{eff}= - U_{am}\cos{(2kx\cos{\alpha})}\,,
\end{equation}
\begin{equation}\label{uam1}
   U_{am}= \frac{e^2A_0^2
   \left( 
     (1+\cos^2{\alpha})\beta_\parallel^2-\cos(2\alpha)- 2\beta_\parallel\sin{\alpha}
   \right)}
   {2 \gamma_\parallel m c^2(1-\beta_\parallel\sin{\alpha})^2 }\,,
\end{equation}
where $\beta_\parallel=v_0/c$ and positive $\beta_\parallel$ means that the
electron moves in the direction of $\mathbf{k}_s=\mathbf{k}_1+\mathbf{k}_2$,
while negative $\beta_\parallel$~-- in opposite direction.

In the case of \emph{equal circular lasers polarization} (for both lasers it is
either right-handed or left-handed) in the same beams geometry the use of the
described method yields in the following effective potential amplitude
expression
\begin{equation}
  \label{uamc}
  U_{am}^c=\frac{e^2 A_0^2 
    (1+2\beta_{\parallel}^2-
    \cos{(2\alpha)}-4\beta_{\parallel}\sin{\alpha})}
  {2\gamma_\parallel mc^2(1-\beta_{\parallel}\sin{\alpha})^2}
\end{equation}
The method is applicable to a system only when both rapid momentum and
trajectory oscillations are considerably less then the slow ones:
$\xi_i\ll\bar{x}_i$ and $p_{i}^{\xi}\ll\bar{p}_i$. The particle trajectory in
such a system is characterized by
$\Omega_1=2k\cos\alpha\sqrt{U_{am}/(\gamma_\parallel m)}$ frequency for
channeling oscillations (within standing wave structure) and
$\Omega_2=\omega_0 (1-\beta_\parallel\sin\alpha)$ frequency for rapid
oscillations due to the particle interaction with the interference laser field.

\subsection{Potential inversion}
\begin{figure}
\includegraphics[width=8.5cm]{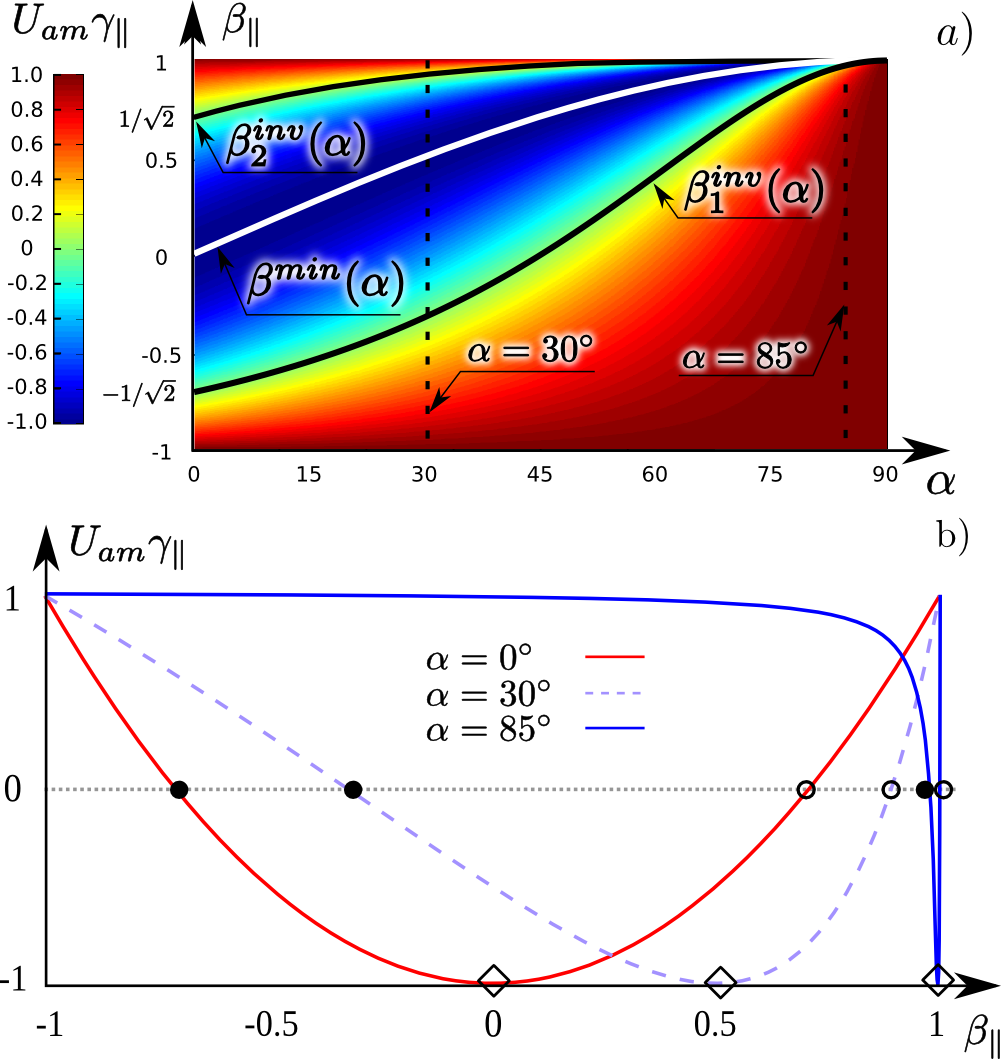}
\caption{\label{f2} a) The averaged potential amplitudes normalized by
  $ (e^2 A_0^2) / (2\gamma_\parallel m c^2)$ as a function of $\beta_\parallel$
  and $\alpha$ is shown in color. Inversion speed values $\beta^{inv}_{1,2}$ as
  a functions of $\alpha$ are shown in black. And $\beta^{min}(\alpha)$ is shown
  in white. b) Normalized potential amplitudes for two p-polarized laser beams,
  crossed at different angles. The inversion speed values $\beta^{inv}_1$ are
  marked with $\bullet$, values of $\beta^{inv}_2$~-- with $\circ$, and
  $\beta^{min}$~-- with $\diamond$ for each $\alpha$.}
\end{figure} 
The above derived expressions (\ref{ueff1}) and (\ref{uam1}) describe the
potential in the region of two p-polarized laser crossing beams. For $\alpha=0$
the beams are counterpropagating and the geometry is similar to that considered
in \cite{kaplanprl,kaplanpra,smorenburg11}. The potential amplitude in this case
is positive for $\left|\beta_\parallel\right|>1/\sqrt{2}$ and negative
otherwise.  This corresponds to the results reported in
\cite{kaplanprl,kaplanpra,smorenburg11} and means that for an electron moving at
the speed
$\left|\beta_\parallel\right|=1/\sqrt{2} =-\beta^{inv}_1=\beta^{inv}_2$
\emph{no} periodic potential is formed. Let us call these two values the
``\emph{inversion speed}'' $\beta^{inv}_{1,2}$. And $\beta^{min}$ is the
electron speed value for which at chosen $\alpha$ the value of
$U_{am}\gamma_\parallel$ becomes minimal. When the electron speed is
$\left|\beta_\parallel\right|>1/\sqrt{2}$, the regions of interference electric
field peaks are scattering for it. And for
$\left|\beta_\parallel\right|<1/\sqrt{2}$ these regions are attracting. This
potential inversion is observed only for p-polarized crossed laser beams.

The expressions (\ref{ueff1}) and (\ref{uam1}) covers arbitrary lasers crossing
angles and the analysis of Eq.~(\ref{uam1}) shows that the values of inversion
speed may be expressed by
\begin{equation}
  \label{eq:bp}
  \beta^{inv}_{1,2}(\alpha)=
  \frac{\sin\alpha\mp\sqrt{2}\cos^2\alpha}{1+\cos^2\alpha}
\end{equation}
This function is presented in Fig.~\ref{f2}.a. One may see that
$\left|\beta^{inv}_1\right|=\left|\beta^{inv}_2\right|$ only for
$\alpha=0$. Varying $\alpha$ causes shift of the
$\left[ \beta^{inv}_1,~\beta^{inv}_2\right]$ range to positive $\beta_\parallel$
region (see Fig.~\ref{f2}.b). This means that an electron with
$\beta_\parallel=0.978$ in the region of two laser beams overlapping at the
angle of $10^\circ$ ($\alpha=85^\circ$) is affected by no ponderomotive force
($U_{am}\gamma_\parallel=0$), whereas an electron with $\beta_\parallel=0.9962$
could become channeled in the effective potential ($U_{am}\gamma_\parallel=-1$).

For the chosen electron longitudinal energy values of lasers crossing angle
$\alpha=\alpha^{min}$, for which
$U_{am}\gamma_\parallel(\alpha^{min},\beta_\parallel)=-1$, could be easily
derived from the expression (\ref{uam1})
\begin{equation}
  \label{bmin}
  \sin\alpha^{min}=\beta_\parallel
\end{equation}
Such electron in the field of two lasers crossed at the angle of $\alpha^{min}$
with oscillates near the transverse electric field peaks. On the other hand,
channeled electron with $\beta_\parallel<\beta^{inv}_1$ (or
$\beta_\parallel>\beta^{inv}_2$) oscillates near the transverse electric field
nodes plane.

\begin{figure}
\includegraphics[width=8.5cm]{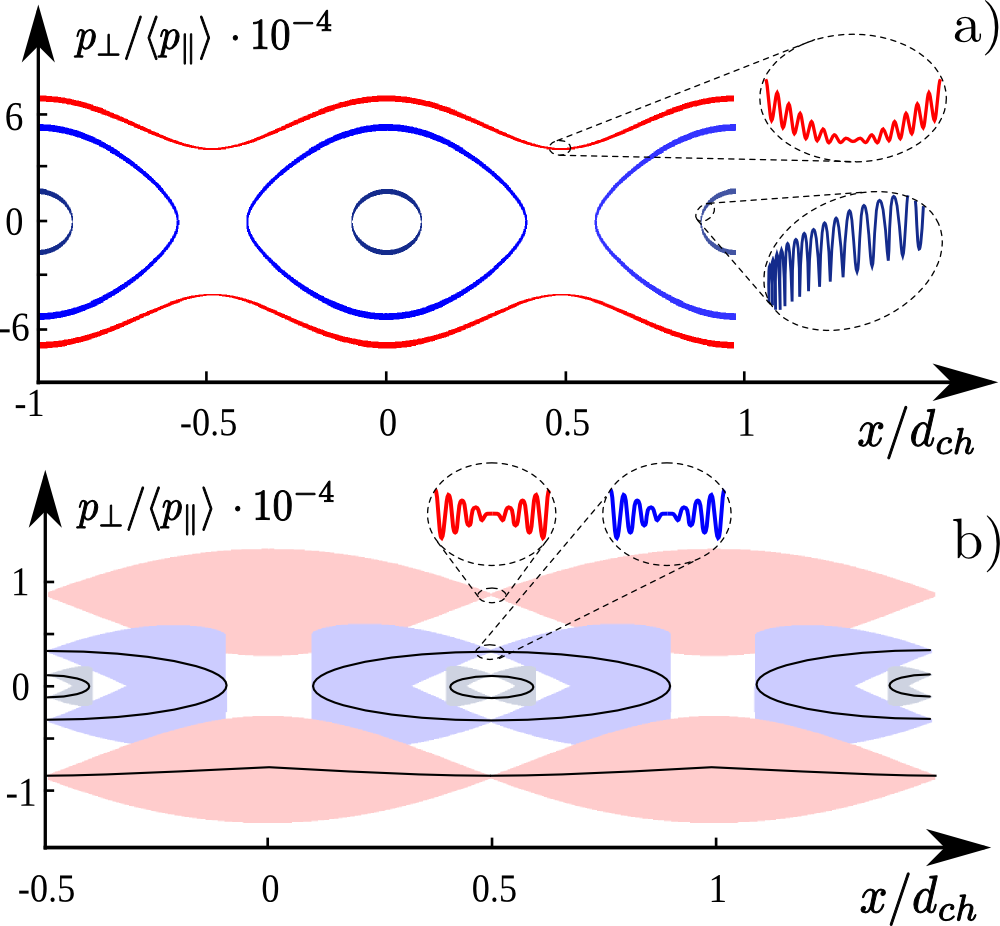}
\caption{\label{f3} Numerically calculated transverse phase-space trajectories
  for electron with $\beta_\parallel=\beta^{min}$ (a) and
  $\beta_\parallel=0.99986$ (b) moving in a channel formed by two p-polarized
  lasers of $\lambda=800~nm$ wavelength and $5~TW/cm^{2}$ intensity for each
  one. The electron transverse momentum is normalized by the averaged
  longitudinal momentum, while the transverse coordinates~-- by the channel
  width $d_{ch}\equiv\lambda_0/(2\sin\eta)$. Channeled (under-barrier)
  trajectories are shown in blue, while quasichanneled (over-barrier)~-- in
  red. The rapid oscillations could not be properly visualized and are outlined
  near the plots. The averaged phase space trajectories are also demonstrated in
  solid black lines. To build these trajectories the 4$^{th}$ order Runge-Kutta
  method was used for the electrons motion equations integration. }
\end{figure}

In Fig.~\ref{f3} the projections of phase space trajectories onto the plane of
transverse momentum and coordinate are presented for both channeled and
over-barrier electrons in the field of laser beams crossed at the angle of
$30^\circ$ ($\alpha=75^\circ$). The inner closed curves (in blue) correspond to
channeled electrons with transverse energies less than potential well height.
Over-barrier electrons (in red) are characterized by transverse energies greater
than the barrier height and not limited within the channels.

For electron with $\beta_\parallel=\beta^{min}$ the channel borders are situated
at $x/d_{ch}=0.5+n$ and, correspondingly, the channel centers~-- at $x/d_{ch}=n$
(see Fig.~\ref{f3},a). On the contrary, electron with longitudinal velocity
$\beta_\parallel\to 1 > \beta^{inv}_2$ could be trapped by the channels, central
axes of which are placed at $x/d_{ch}=0.5+n$ (see Fig.~\ref{f3},b), where
$n=0, \pm 1, \pm 2, \dots$

\begin{figure}
\includegraphics[width=8.5cm]{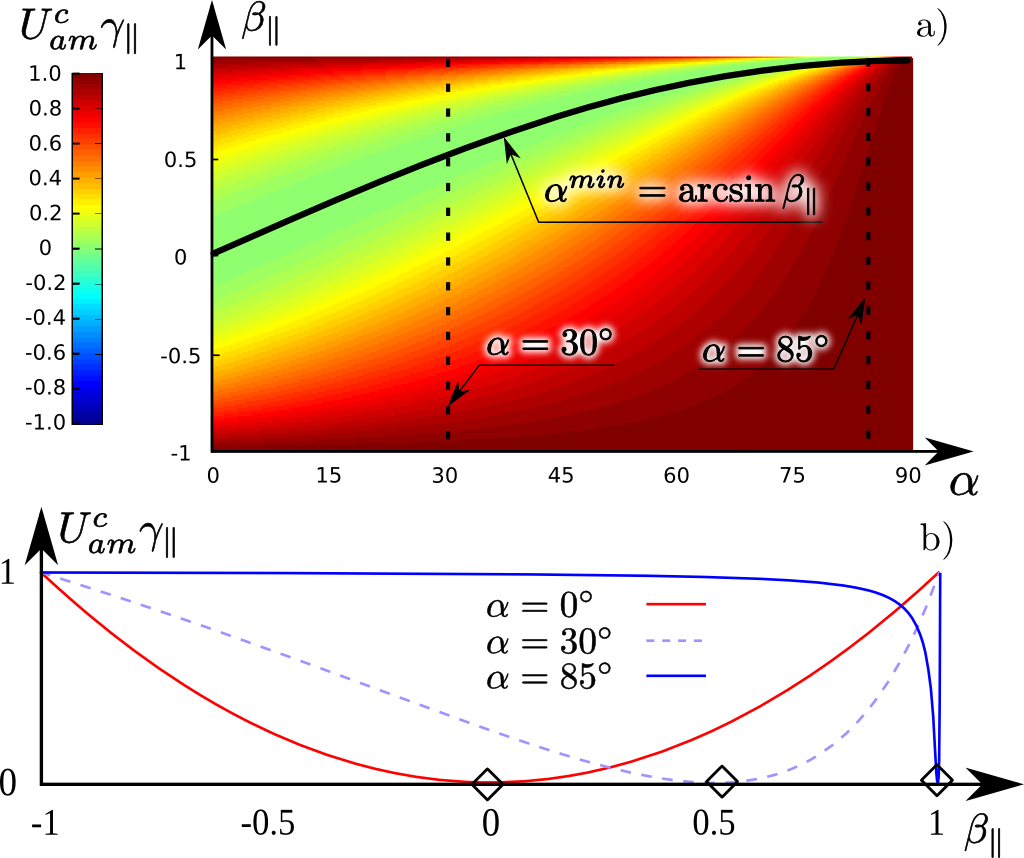}
\caption{\label{f4} a) Normalized to unity potential amplitude
  $U_{am}^c\gamma_\parallel$ as a function of $\beta_{\parallel}$ and $\alpha$
  for circularly polarized lasers case is shown. The amplitude of effective
  potential reaches minimal value equal zero on the curve
  $\alpha^{min}=\arcsin\beta_{\parallel}$.  b) Normalized potential amplitudes
  for two circularly polarized laser beams, crossed at different angles. The
  values $\beta^{min}=\sin\alpha$ are marked with $\diamond$ for each $\alpha$.}
\end{figure}

Notably, the effective potential $U^c_{am}\gamma_\parallel$ for the case of
equal circular lasers polarization normalized to unity has the form exactly
similar to Eq.~(\ref{uam1}) but its minimal value is zero (see
Fig.~\ref{f4}). This case for $\alpha=0$ was considered in
\cite{kaplanprl,kaplanpra,smorenburg11} previously, providing the same
results. But the expression (\ref{uamc}) describes the potential amplitude for
arbitrary $\alpha$ showing that an electron in the field of circularly polarized
lasers crossed at the angle of $\alpha^{min}=\arcsin\beta_\parallel$ feels no
effective potential, and no inversion is observed for circularly polarized laser
beams. This is the main difference comparing to the p-polarization case.

For details, in the supplemental materials \cite{supp} one can find two
animations of channeled electron dynamics numerical simulations. These results
were received by motion equations integration for the electron in the field of
p-polarized and circularly polarized electromagnetic waves. The 4$^{th}$ order
Runge-Kutta method was used. The projectile longitudinal velocity varies from
$\beta_\parallel=-0.969$ to $\beta_\parallel=0.999$ just as if the additional
longitudinal electric field was imposed on the region. Current electron
longitudinal speed is shown in the upper region of the video. Spatial
distribution of the normalized potential distribution
$U_{am}\gamma_\parallel(x/d_{ch})$ is shown for every moment of time in the
right region of the video. In the main area averaged transverse oscillations of
the electron is put onto the current potential distribution presented in color.
\section{Channeled electrons radiation}
Obviously, the possibility of charged beam channeling in electromagnetic
lattices becomes rather interesting due to the ability of beams shaping by means
of tools based on this process. Moreover, speaking on possible applications of
the interaction under consideration one should outline that it can be used as a
promising tunable source of intense radiation. The radiation spectrum of a
charged particle emitting in a crossed laser field is characterized by two peak
frequencies $\omega_1=\Omega_1/(1-\beta_\parallel\cos(\theta))$ and
$\omega_2=\Omega_2/(1-\beta_\parallel\cos(\theta))$ \cite{rreps}, where
$\Omega_{1,2}$ were defined before. Both of them being measured in the forward
direction ($\theta=0$) are shifted by the factor of $\sim2\gamma_\parallel^2$
due to the Doppler effect. The first one $\omega_1$ corresponding to slow
channeling electron oscillations in the potential well of a system does not
depend on the laser frequency $\omega_0$.  The $\omega_2$ radiation frequency is
caused by the electron interaction with the interference laser beams
field. Hence, $\omega_2$ is strictly defined by the electron velocity as well as
by the wave vectors $\mathbf{k}_1$ and $\mathbf{k}_2$.  The intensity of
electromagnetic radiation by a laser-channeled electron can be described by the
4-potential $A_{\mu}=(\mathbf{A},\varphi)$
\begin{equation}
  \label{eq117.1}A_{\mu}(\mathbf{r},t)=
  \frac{4\pi}{c}\int j_{\mu}(\mathbf{r}',t')
  G(\mathbf{r}-\mathbf{r}',t-t')d^3r'dt'\,,
\end{equation}
where $j_{\mu}=(\mathbf{j},c\rho)$ is the 4-current,
$G(\mathbf{r}-\mathbf{r}',t-t')$ is the Green function. The field can be found
in far-field zone and represented as a sum of spherical monochromatic waves.
Taking into account that relativistic particle emits radiation in a narrow
forward directed solid cone, one could define the analytical expression for
radiation spectrum of relativistic channeled electron moving near the bottom
(center) of the cross-laser channel \cite{rreps}
\begin{equation}\label{eq117}
  \frac{dP}{d\omega}=\sum \limits_{i=1}^2 
  \frac{e^2\Omega_i^3a_i^2\gamma_\parallel^2}{c^3}
  \zeta_i\left(1-2\zeta_i+2\zeta_i^2\right)\Theta\left(\pi N_i(1-\zeta_i)\right)\,,
\end{equation}
where $\omega$ is the radiation frequency, $a_1$ is the amplitude of channeling
oscillations, and $a_2$ is the amplitude of rapid oscillations, $N_i$ is the
number of $\Omega_i$-frequency particle oscillations,
$\zeta_{i}=\omega/(2\gamma_\parallel^2\Omega_i)$, and
$\Theta(x)=0.5+\mathrm{si}(2x)/\pi-\sin^2(x)/\pi x$ \cite{bag02,kum91}. The
results of numerical computations of the electron radiation emitted in the
considered system are similar to well known crystal channeling radiation spectra
(e.g. see spectra in \cite{rreps} and \cite{bog08}).

Based on previously derived expressions, simple estimations of a single electron
radiation yield are presented. For the external laser intensities
$I\sim10^{14}~W\,cm^{-2}$ the maximum of emitted channeling radiation spectral
distribution falls on $\lambda_1\approx1~\mu m$ for a channeled electron with
$\gamma_\parallel\approx10^3$. And for the external laser wavelength of
$\lambda_0=1~\mu m$ crossed at $\eta=15^\circ$ the channels width would be
$d_{ch}\equiv\lambda_0/(2\sin(\eta))\approx2\lambda_0\approx2~\mu m$. Total
power emitted by the channeled electron can be then evaluated by
\begin{equation}
  P[W]=\frac{2e^2\Omega_1^4a_1^2\gamma_\parallel^4}{3c^3}\approx
  10^{-42}\left(I\left[\tfrac{W}{cm^2}\right]\right)^2,
\end{equation}
for the channeling oscillations amplitude $a_1\approx0.1d_{ch}$.

One should note that, first, since both electron and laser parameters are chosen
to meet parametric resonance conditions \cite{Akh91} the radiation wavelength is
equal to the external laser one. Hence, this could be interesting as radiation
amplification mechanism. Moreover, due to rather wide channel width it becomes
possible to use a large bunch of, for instance, $10^{10}$ electrons, radiating
coherently with a power
\begin{equation}
P[W]\approx 10^{-22}\left(I\left[\tfrac{W}{cm^2}\right]\right)^2
\end{equation}
As seen, in the case of initial laser intensity of $10^{17}~W\,cm^{-2}$ the
total radiation power is estimated to be equal to $10^{12}~W$.

Obviously, there are some constraints imposed on channeled electrons.  One of
them is the bunch divergence. Its critical value could be derived from the
channel potential height, and for the parameters above used it can be estimated
as $\sim 0.1~mrad$ (it corresponds to the values of
$p_\perp/\langle p_\parallel\rangle$ shown in Fig.~\ref{f3} for over-barrier
electrons). In addition, there is an upper limit for the laser intensities to
form the channel structure. The limit is caused by the fact that for some laser
intensities the rapid electron oscillations become too high allowing the
electron to leave the channel independently of its initial transverse velocity
(electron dechanneling). This is caused by system transition to chaotic behavior
described in \cite{steeb95}. Kapitza method is no applicable for such regimes
which could be observed for different crossing angles as well as laser
polarizations. But their investigation seems to be very promising for
applications as a future new-type $\gamma$-radiation source when needed
intensities will become achievable (see \cite{art14} for details).

\section{Summary}
Summing up, channeling of an electron in the field of crossed laser beams was
reconsidered with a special attention paid to the peculiarities of the
interaction potential. The potential derived in terms of classical physics
reveals strong dependence on external lasers parameters and their mutual
orientation. Moreover, it shows rather complex dependence on the electron
longitudinal velocity by its value and direction. For the case of p-polarized
lasers, the increase in channeled electron longitudinal velocity from
$\beta_\parallel\to-1$ (which corresponds to the opposite to $Oz$-axis electron
motion) to $\beta^{inv}_{1}$ results in the decrease of the potential amplitude
$U_{am}\gamma_\parallel$ down to zero. In the $[\beta^{inv}_{1};~\beta^{min}]$
interval the potential becomes scattering and grows to its maximum at
$\beta^{min}$. At the following increase of $\beta_\parallel$ the scattering
potential fades away to zero at $\beta^{inv}_{2}$. For $[\beta^{inv}_{2};~1)$
the potential becomes attracting again, growing from zero to its maximum at the
end of the interval (this process is shown in \cite{supp}).

The averaged potential $U_{am}^c\gamma_\parallel(\alpha,\beta_\parallel)$ for
the laser beams of equal circular polarization has exactly the same shape as in
the case of p-polarized beams. But its minimal value lying on the line
$\beta^{min}=\sin\alpha^{min}$ is zero, hence, no potential inversion is
observed.

The radiation spectrum classically calculated is characterized by two major
frequencies: the first one is due to external laser field scattering on
relativistic electron, while the second corresponds to the radiation of electron
trapped by the effective potential well. The maximum radiation intensity falls
on the frequency $\omega_m^{rad}\sim \gamma_\parallel\sqrt{I}$ defined by both
electron energy and external laser intensity, while the integral radiated power
depends on squared external laser intensity $P\sim I^2$.  Combined with high
radiation coherence for a channeled electron bunch it can result in a high
intensity gain that needs more detailed investigation.

~

This work was supported by the Ministry of Education and Science of RF in the
frames of Competitiveness Growth Program of National Research Nuclear University
MEPhI, Agreement 02.A03.21.0005 and the Dept. for Education and Science of LPI
RAS.

\bibliography{prl-dabagov-dik-frolov}
\end{document}